**Updated determination of the molar gas constant *R* by acoustic measurements in argon at UVa-CEM.**

Segovia J.J. [a], Lozano-Martín D. [a], Martín M.C. [a], Chamorro C.R. [a], Villamañán M.A. [a], Pérez E. [a], García Izquierdo C. [b], del Campo D. [b]

[a] TERMOCAL Research Group University of Valladolid (UVa), Paseo del Cauce 59, 47011, Valladolid, Spain.

[b] CEM (Centro Español de Metrología), Alfar 2, 28760 Tres Cantos, Madrid, Spain.

* To whom correspondence should be addressed (e-mail): jose.segovia@eii.uva.es

**Abstract**

A new determination of the molar gas constant was performed from the measurements of the speed of sound in argon at the triple point of water and extrapolation to zero pressure. A new resonant cavity was used. This is a triaxial ellipsoid whose walls are gold-coated steel, and which is divided into two identical halves that are bolted and sealed with an O-ring. Microwave and electroacoustic traducers are located in the northern and southern parts of the cavity, respectively, so that measurements of microwave and acoustic frequencies are carried out in the same experiment. Measurements were taken at pressures from 600 kPa to 60 kPa and at 273.16 K. The internal equivalent radius of the cavity was accurately determined by microwave measurements and the first four radial symmetric acoustic modes were simultaneously measured and used to calculate the speeds of sound. The improvements made using the new cavity have reduced by half the main contributions to the uncertainty due to the radius determination using microwave measurements which amounts to 5.1 parts in $10^6$ and the acoustic measurements, 4.4 parts in $10^6$, where the main contribution (3.7 parts in $10^6$) is the relative excess half-widths associated with the limit of our acoustic model, compared with our previous measurements. As a result of all the improvements with the new cavity and the measurements performed, we determined the molar gas constant $R = (8.314449 \pm 0.000059)$ J·K$^{-1}$·mol$^{-1}$ which means a relative standard uncertainty of 7.0 parts in $10^6$. Although this uncertainty is greater than the most recent determinations of the molar gas constant, the value reported in this paper lies -1.4 parts in $10^6$ below the recommended value of 2014 CODATA, although still within the range consistent with it.

**1. Introduction.**

The important request of the General Conference on Weights and Measures (CGPM) and the International Committee for Weights and Measures (CIPM) to the Committee on Data for Science and Technology (CODATA) of a special adjustment to determine recommended values of universal constants for the revised definition of the International System of Units (SI) [1] has motivated the present work. This paper provides an update for the determination of molar gas constant $R$ previously reported by our group [2] through the extrapolation to zero pressure of speed of sound measurements in argon at the temperature of the triple point of water, $T_{TPW} = 273.16$ K. As stated in our previous report [2], the greatest contribution to the relative standard uncertainty of 16 parts in $10^6$ in the Boltzmann constant $k_B$ arose from the microwave cavity radius determination. To overcome this limitation, a new acoustic resonance cavity was acquired based on a stainless-steel triaxial ellipsoid coated in gold, instead of the misaligned bare stainless-steel sphere used before. Also, some new contributions to the resonance frequency corrections in our acoustic model have been considered, following the data reduction and fitting process of [3] - [4]. The experimental details are described in section 2 and the results of the experiments are shown in section 3 along with a discussion of our new estimate for $R$ and its uncertainty.

## 2. Experimental set-up.

### 2.1 Quasi spherical cavity description and geometrical characterization.

The new resonance cavity has been designed according to the instructions given in [5] and made by the mechanical department of DG-Technology Service s.r.l. in austenitic stainless-steel creep-free of grade 316L. It is coated internally, with a 15 μm film of gold using a plasma treatment unit as part of a vacuum coater system. Its geometrical shape, described by equation 1, is that of a triaxial ellipsoid with axes of length $a$, $a \cdot (1+\varepsilon_1)$; $a \cdot (1+\varepsilon_2)$:

$$\frac{x^2}{a^2} + \frac{y^2}{a^2(1+\varepsilon_1)^2} + \frac{z^2}{a^2(1+\varepsilon_2)^2} = 1 \qquad (1)$$

where $a = 40$ mm, $\varepsilon_1 = 0.002$, $\varepsilon_2 = 0.001$ are the nominal internal radius, and eccentricities specified to the manufacture, respectively. A schematic diagram of the cavity is displayed in figure 1. It is made from two quasi-hemispheres bolted together with 12 M6 bolts tightened to 10 Nm of torque and sealed with O-ring of perfluoroelastomer (Kalrez). Two equal acoustic transducers, source and detector, are placed in the southern hemisphere at the south pole ($\theta_s$) = 0 and ($\theta_d$, $\phi_d$) = (cos$^{-1}$ ($\sqrt{3/5}$), $\pi/4$), with their diaphragm flush with the inner surface of the shell (the co-ordinate system is indicated in figure 1). They are non-commercial capacitance type microphones based on Professor J.P.M. Trusler's designs. Details of their construction are shown elsewhere [6] and a sketch of our units is shown in [7]. The northern hemisphere is provided with two antennas at positions ($\theta_s$, $\phi_s$) = ($\pi/4$, $\pi/4$) and ($\theta_d$, $\phi_d$) = ($\pi/4$, $5\pi/4$). They are made of a welded oxygen-free high conductivity copper cable with triple-loop shape nearly 1 mm of diameter to couple to both transverse magnetic, TM, and transverse electric, TE, microwave modes, they are described in [7]. The loop cable is flush with the inner surface of the shell in order to contribute as little as possible to acoustic half-width and epoxy resin fills the plugs to provide a pressure-tight shell.

As described in [8], [9] the internal equivalent radius $a_{eq}$ of a perfect spherical cavity with the same volume as our quasi-spherical cavity is given by:

$$a_{eq} = a[(1+\varepsilon_1)(1+\varepsilon_2)]^{\frac{1}{3}} \qquad (2)$$

and could be accurately determined by microwave measurement of the components of triply degenerate TE$_{1n}$ and TM$_{1n}$ modes due to the different ellipsoid axes length as:

$$\langle f_{1n} \rangle = \frac{1}{3}(f_x + f_y + f_z) = \frac{z_{1n} c}{2\pi a_{eq}} \tag{3}$$

where $f_m$ with $m = x, y, z$ are three components of the triplets, $\langle f_{1n} \rangle$ is the average resonance frequency, $c$ is the speed of light and $z_{1n}$ are the eigenvalues of the Helmholtz equation that can be numerically calculated from the zeros of the spherical Bessel function $j_l$ for TE modes and the zeros of $[j_l(z)+zj_l'(z)]$ function for TM modes. We have performed the geometrical characterization by means of the first three TM$_{1n}$ modes and the first three TE$_{1n}$ modes, getting three well-resolved components in all the cases thanks to the high electrical conductivity of the gold coating. This significantly reduces the resonance half-widths, compared with the broad microwave doublets previously measured with the misaligned cavity [2]. An Agilent Network Analyzer VNA N5230C, calibrated with an Agilent Calibration Kit 8510, is configured to measure the complex scattering coefficient $S_{21}$ through the cavity and connected to the antennas with GoldPt SMB r/a plg-plg RG316 waveguides. The timebase of the VNA is linked to a rubidium standard frequency to improve the accuracy and the thermal stability of microwave frequency measurements. LabVIEW software originally programmed by Professor Eric May controls the microwave system. The network analyzer was set to sweep through 201 discreet frequencies; they are the average of five scans with a frequency bandwidth IFBW of 10 Hz and measuring simultaneously with the acoustic determinations. The software fits the real $u$ and imaginary $v$ signal by an iterated procedure depending on the sum of three Lorentzian functions:

$$u + iv = \sum_{m=x,y,z} \frac{2if g_{1n}^m A_{1n}^m}{(F_{1n}^m)^2 - f^2} + B + C(F_{1n}^m - f) \tag{4}$$

where $A$, $B$, and $C$ are complex constants and $F_N = f_N + ig_N$ are the complex resonance frequencies. The data are first corrected by the skin effect [8] that takes into account the finite electrical conductivity of the boundary wall layer of gold:

$$\left(\frac{\Delta f_{skin} + ig}{f}\right)_{1n} = \frac{\delta}{2a}(-1 + i) \quad \text{for TE modes} \tag{5}$$

$$\left(\frac{\Delta f_{skin}+ig}{f}\right)_{1n} = \frac{\delta}{2a}(-1+i)\frac{z_{1n}^2}{z_{1n}^2-2} \quad \text{for TM modes}$$

where $\delta = (\pi\mu\sigma f)^{-1/2}$, $\mu(p,T)$ is the magnetic permeability (in our case, $\mu(p,T) = \mu_0 = 4\pi 10^{-7}$ N/A$^2$ in vacuum) and $\sigma$ is the electrical conductivity of the wall material, determined from the experimental microwave half-width $g$. We use our experimental half-widths as the frequency corrections for the skin effect and the skin depths ($\delta$) were obtained using equation (5), the values range from $1.6 \cdot 10^{-6}$ m for TE$_{13}$ mode to $2.5 \cdot 10^{-6}$ m for TE$_{11}$ mode, the average skin depth was used to estimate the value of the electrical conductivity $\sigma_{exp} = 7.6\ 10^6$ S m$^{-1}$, this calculation procedure was also used in [10]. This value is 6.4 smaller than the value found in the literature $\sigma_{lit} = 4.88\ 10^7$ S m$^{-1}$ [11] for pure gold at $T_{TPW}$. This difference possibly comes out from the discontinuities in the wall due to the gap at the equatorial junction between the hemispheres and the slits around the transducer and antenna plugs, because as can be seen in figure 2, the microwave excess half-widths are greater for those modes that have higher density currents in the wall surface, like TM$_{11}$ and TE$_{11}$ modes.

The average resonance frequency of the triplet is corrected by the geometrical perturbation due to the gas inlet duct [12] of radius $r = 0.5$ mm. This correction is independent of the hole position on the wall and also on the duct length for long tubes:

$$\langle\frac{\Delta f_{duct}}{f}\rangle_{1n} = -\frac{r^3}{4\pi a^3}\frac{0.950 z_{1n}^2 - 1.152}{z_{1n}^2 - 2} \quad \text{for TM modes}$$

$$\langle\frac{\Delta f_{duct}}{f}\rangle_{1n} = -\frac{r^3}{4\pi a^3} 0.950 \quad \text{for TE modes} \tag{6}$$

No correction was applied for the perturbation of the antennas because no complete theory to model loop antennas is available in the literature. In any case, the improvement in the agreement between modes using the theory of straight antennas [12] results negligible perturbations in our case. The eccentricities are determined from the first order perturbation theory predictions of the fractional frequency shifting of the triplets components [5] due to the non-sphericity of the cavity:

$$\frac{f_m - f_0}{f_0} = \frac{2}{15}\left(-\frac{1}{2} - \frac{3}{z_{1n}^2 - 2}\right)\begin{cases}(-2\varepsilon_1 + \varepsilon_2)\ m = x \\ (\varepsilon_1 + \varepsilon_2)\ m = y \\ (\varepsilon_1 - 2\varepsilon_2)\ m = z\end{cases} \text{for TM modes} \tag{7}$$

$$\frac{f_m-f_0}{f_0} = \frac{2}{15}\left(-\frac{1}{2}\right)\begin{cases}(-2\varepsilon_1+\varepsilon_2)\ m=x\\(\varepsilon_1+\varepsilon_2)\ m=y\\(\varepsilon_1-2\varepsilon_2)\ m=z\end{cases}\text{ for TE modes}$$

where $f_0 = \langle f_{1n}\rangle = (f_x+f_y+f_z)/3$ is the average triplet frequency. The experimental average values of $\varepsilon_1$ and $\varepsilon_2$ determined from all modes of our microwave measurements are $\varepsilon_1 = 0.00215$ and $\varepsilon_2 = 0.00119$, with an experimental standard deviation of the mean, $\sigma(\bar{\varepsilon})=\pm 0.00003$, and close to those specified to the manufacturer $\varepsilon_1 = 0.002$ and $\varepsilon_2 = 0.001$. These results are used to correct the average microwave frequency for the second order shape perturbation [13], [14]:

$$\langle\frac{f_m^2-f_0^2}{f_0^2}\rangle_{1n} = \frac{2(33z_{1n}^8-245z_{1n}^6+714z_{1n}^4-1152z_{1n}^2+160)}{1125(z_{1n}^2-2)^3}(\varepsilon_1^2-\varepsilon_1\varepsilon_2+\varepsilon_2^2)$$
for TM modes

$$\langle\frac{f_m^2-f_0^2}{f_0^2}\rangle_{1n} = \left(\frac{22z_{1n}^2}{375}-\frac{2}{225}\right)(\varepsilon_1^2-\varepsilon_1\varepsilon_2+\varepsilon_2^2)\text{ for TE modes}$$

(8)

Using the values of the relative electrical permittivity $\varepsilon_r(p,T)$ for argon from REFPROP 9.1 (updated to the last version at June 10, 2014) [15] the corrected resonance frequencies are set to the equivalent values of the triple point of water and average pressure through all repetitions by:

$$\langle f_{1n}(T_{\text{TPW}},p_{\text{avg}})\rangle = \langle f_{1n}(T_{\text{exp}},p)\rangle\left(1+\alpha_T(T-T_{\text{TPW}})\right)\left(1+\frac{\kappa_T}{3}(p-p_{\text{avg}})\right) \quad (9)$$

where $\kappa_T$ [16] and $\alpha_T$ [17] are isothermal compressibility and thermal expansion coefficient of stainless steel of grade 316L, respectively, and $T_{\text{exp}}$ is close to the temperature of the triple point of water. The equivalent internal radius of the quasi-spherical cavity is then calculated by:

$$a_{eq} = \frac{z_{1n}c}{2\pi(\langle f_{1n}\rangle - \Delta f)} \quad (10)$$

where $\Delta f = \Delta f_{\text{skin}} + \Delta f_{\text{duct}} + \Delta f_{\text{shape}}$ and $c = 1/(\mu(p,T)\varepsilon_r(p,T)\cdot\varepsilon_0)^{1/2}$, $\mu(p,T) = \mu_0 = 4\pi 10^{-7}$ N/A$^2$ and $\varepsilon_0 = 1/(35950207149\pi)$ F/m in vacuum. Table 1 shows the measured radius at each pressure simultaneously with the acoustic measurements at $T_{\text{TPW}}$. The TM$_{11}$ mode is discarded from the average of the radius because of its value is not in agreement with the other modes [2]. As can be seen in table 1, the radius is increased with the pressure. The values obtained for the equivalent internal radius at zero pressure of the different modes are shown in Table 2. The scatter of the zero pressure radius values for the different modes is used to estimate one of the contributions of the radius uncertainty to the molar gas constant $R$, the experimental standard deviation of the mean of the zero pressure radius is $8.65\cdot 10^{-8}$ m, which means a relative standard uncertainty of 2.1 parts in $10^6$ to radius, which amounts to 4.3 parts in $10^6$ to $R$. There is a second term in the uncertainty of the radius which arises from the difference between the theoretical and experimental half-widths, the mean of excess half-widths is 1.3 parts in $10^6$, and it is included in the uncertainty budget estimating a relative standard uncertainty of 2.7 parts in $10^6$ to $R$.

Additionally, from a linear fit of radius versus pressure an experimental value of the isothermal compressibility of the shell $\kappa_{T,exp} = 2.60 \cdot 10^{-11}$ Pa$^{-1}$ was calculated, which is one order of magnitude higher than the value of the literature $\kappa_{T,lit} = 5.60 \cdot 10^{-12}$ Pa$^{-1}$ for 316 L stainless steel [16]. We explain this deviation of nearly one order of magnitude due to our configuration, in which the pressure is only changing inside the cavity against a high vacuum in the outside and the radius is increased with the pressure. We have estimated an uncertainty contribution due to this linear fit of 0.2 parts in $10^6$ to $R$.

**2.2 Pressure and temperature measurement.**

Pressure and temperature measurement system are extensively described in our previous papers [18], [19], so this section is only a summary. Pressure is determined with a piezoelectric Digiquartz 2300A-101 transducer thermally isolated at the top of the gas inlet tube. The effect of the hydrostatic column of gas is added to the pressure measurement. The pressure gauge has been calibrated up to 2 MPa prior the experiment in our facilities with a dead weight tester (DH 5213) with load masses calibrated with international traceability in the Spanish National Institute (Centro Español de Metrología -CEM) laboratories. The standard uncertainty in pressure is $u(p) = 3.75 \cdot 10^{-5}$ ($p$/Pa) + 100 Pa. This contribution of calibration is 0.14 parts in $10^6$ to $R$, when extrapolating at zero pressure the speed of sound. The repeatability during the experiment is better than this value but we have assigned the same amount, therefore, the pressure measurement gives a contribution to the relative uncertainty of $R$ of 0.2 parts in $10^6$.

Temperature is determined as the mean of two readings from Inconel X-750 capsule-type standard platinum resistance thermometers CSPRT Rosemount 162D of 25.5 Ω, located at north and south quasi-hemispheres and plugged in four wire configuration to an ac bridge

Automatic Systems Laboratories ASL F18, which is referenced to 25 Ω thermostated resistance Tinsley 5685A. A maximum difference of 0.5 mK has been observed between north and south probes in this work. The external standard resistance has been calibrated in CEM laboratories, but its calibration uncertainty does not affect the uncertainty of temperature measurement. The bridge has been configured to get the maximum accuracy in temperature measurement: source impedance set to 10 Ω, 1mA carrier at 75 Hz, in-phase detector gain to $10^4$ and bandwidth to 0.1 Hz in automatic balance mode. The CSPRTs have been calibrated again before this experiment at $T_{TPW}$ by using a triple point of water cell previously compared with the group of cells which maintain the Spanish national standards following the procedure of the triple point of water fixed-point realization reported in [19]. The self-heating effect has been taken into account using the procedure described in [7] using the same measurement bridge, reference resistance and connection cables as later. A drift of 0.1 mK has been observed after the calibration of the CSPRTs, in agreement with the histogram of these probes. We have obtained a standard uncertainty in temperature due to calibration of 0.1 mK. The thermal stability is achieved through three stages of thermostatic control consisting of an exterior aluminium shell and an interior steel jacket, inside the acoustic cavity where all the connection cables are located. Radiation heat loss is minimized by surrounding the jacket by aluminium over fiberglass foils. Convection is eliminated making vacuum inside the jacket with a centrifugal and a turbomolecular pump in series. Temperatures are stabilised by PID controlled resistors attached to the copper block that suspends the acoustic cavity inside and which are also positioned along the side and base of the jacket (see [19] for a detailed description). The stability of temperature through all repetitions amounts to a maximum of 0.6 mK. A detail study of the behaviour of these thermometers was described in [7], all these contributions gives an estimated relative standard uncertainty of 0.9 parts in $10^6$ on $R$.

**2.3 Gas sample details.**

The gas sample handling system consists of an Air Liquid Alphagaz 2 argon of molar purity $x_{Ar}$ = 0.999999 that flows the gas through two oxygen and moisture Agilent Gas Clean

Filter designed for gas chromatography, plus an Agilent Big Universal Trap specific for argon purification, before entering the resonance cavity. This system is similar to our previous Boltzmann constant work [2], the only difference is the inclusion of a bigger capacity universal trap. Then, the uncertainty due to the impurities that changes the mean molar mass of the sample is equal to our past experiment. The supplier states for an argon research grade gas source the following bounds in purity of molar fraction $x$ using their filters: $x(O_2) \leqslant 5 \cdot 10^{-8}$, $x(H_2O) \leqslant 2 \cdot 10^{-7}$, $x(N_2) \leqslant 1 \cdot 10^{-7}$, $x(CO_2) \leqslant 1 \cdot 10^{-7}$, $x(CO) \leqslant 1 \cdot 10^{-7}$, $x(C_mH_n) \leqslant 1 \cdot 10^{-7}$, $x(H_2) \leqslant 2 \cdot 10^{-7}$. Based on this, the amount of non-noble gases and moisture of our gas should be not significant after passing through all the filters. Moreover, we have not observed any effect of progressive contamination as assessed by repeatability tests, despite the non-flow condition of our setup. Noble gases cannot be filtered out with this configuration so in the absence of a chemical analysis, we have taken an upper bound of the effect of impurities based on the thorough studies of the chemical composition of argon performed by other authors with the same gas supplier [3], [4]. In the worst-case scenario there might be amount fractions $x = 2 \cdot 10^{-6}$ mol·mol$^{-1}$ of helium and $x = 2 \cdot 10^{-6}$ mol·mol$^{-1}$ of neon, with negligible krypton and xenon, may result in a change of the mean molar mass of argon $M_{Ar}$ of 2.80 parts in $10^6$. We assume that the probability of having impurities is the same in all the range (0 to 2.8 ppm), this rectangular probability distribution gives a relative standard uncertainty of $R$ of 0.8 parts in $10^6$.

Determination of the isotopic ratios of the three stable isotopes of argon in our sample has not been possible. However, considering that argon is made by an air liquefaction process a representative value $M_{Ar}$ = 0.039947798 kg·mol$^{-1}$ has been assumed from standard isotope abundances $^{40}$Ar, $^{38}$Ar, and $^{36}$Ar in atmospheric air determined in table 7 in reference [20] and atomic weights $M_{40}$, $M_{38}$, and $M_{36}$ given in table 2 in reference [21]. To calculate the isotopic uncertainty in our value of the molar mass, two contributions have been considered: one from the uncertainties of the isotopic ratios stated in [20] and another from the variability of these ratios based on the studies reported in the literature. The uncertainty due to the isotopic ratio determination of argon is given by:

$$u(M_{\text{Ar}})^2 = \frac{1}{\left(1 + \left(\frac{^{36}\text{Ar}}{^{40}\text{Ar}}\right) + \left(\frac{^{38}\text{Ar}}{^{40}\text{Ar}}\right)\right)^4}$$

$$\cdot \left(\left(M_{40} - M_{38} - M_{38}\left(\frac{^{36}\text{Ar}}{^{40}\text{Ar}}\right) + M_{36}\left(\frac{^{36}\text{Ar}}{^{40}\text{Ar}}\right)\right)^2 u\left(\frac{^{38}\text{Ar}}{^{40}\text{Ar}}\right)^2 \right. \tag{11}$$

$$\left. + \left(M_{40} - M_{36} - M_{36}\left(\frac{^{38}\text{Ar}}{^{40}\text{Ar}}\right) + M_{38}\left(\frac{^{38}\text{Ar}}{^{40}\text{Ar}}\right)\right)^2 u\left(\frac{^{36}\text{Ar}}{^{40}\text{Ar}}\right)^2\right)$$

with a relative standard uncertainty of 0.35 parts in $10^6$. Our uncertainty determination from the variability of argon isotopic composition relies on the difference between our mean molar mass estimation and that determined by representative acoustic thermometry works with argon [3]-[4], [22]- [23]. The latest comparisons of isotopic measurements by the Institute for Reference Materials and Measurements, IRMM, and Korean Research Institute of Standards and Science, KRISS, have been taken into account to update this value [21], [24], [25]. An upper limit of 2 parts in $10^6$ obtained from these differences gives a reliable value of the variability of isotopic composition through typical research grade bottles. Thus, a relative standard uncertainty of 1.2 parts in $10^6$ based on a rectangular probability distribution is added in quadrature to yield a combined uncertainty of our estimate of the molar mass of 1.5 parts in $10^6$.

**2.4 Acoustic resonance determination.**

The acoustic dataset consists of a total of 88 measurements taken through the isotherm at $T_{\text{TPW}}$ for 11 pressures values from $p$ = 600 kPa to 60 kPa. The (0,2), (0,3), (0,4), (0,5) radial symmetric acoustical modes were measured simultaneously with the radius measurements by microwave resonance spectroscopy. The experimental acoustics set up is the same as the previous work [2], but for the sake of completeness a description is given. A HP 3225B synthesiser drives the source transducer with a sinusoidal signal raised by a preamplifier to 180 V RMS. The time base of the synthesiser is connected to the standard external frequency of a rubidium clock to reduce the relative standard uncertainty to $10^{-11} \cdot f$. A Lock-In Amplifier SR 850 DSP with its reference connected to the synthesiser detects the response transducer signal at twice the frequency, the second harmonic of the synthesiser, to avoid the crosstalk effect. The detector transducer is powered by a 90 V dc voltage, and plugged through a triaxial cable to an impedance adapter, which buffers the signal to the input of the Lock-In. With the triaxial cable in active guard configuration, the division of the output signal of the small transducer capacitance by the large capacitance of the connections cables is avoided. The system is controlled by a fitting software written in Agilent VEE that takes the real $u$ and imaginary $v$ parts of the signal measure by the Lock-In to perform a Lorentzian fit of 44 points around the resonance frequency $F_{l=0,n} = f_{l=0,n} + g_{l=0,n}$ of the (0,n) mode:

$$u + iv = \frac{if A_{(0,n)}}{(f^2 - F_{(0,n)}^2)} + B + C(f - f_{(0,n)}) \tag{12}$$

where $A$, $B$, and $C$ are complex parameters.

The analysis of the acoustic data to obtain a value for the molar gas constant is based in the acoustic corrective model in references [26] [27] and data processing in references [3], [4], and [28]. The necessary frequency corrections $\Delta f$ to the ideal equation of a perfectly spherical zero wall admittance cavity have been implemented in a similar spreadsheet to that of the previous work [2]:

$$w = \frac{2\pi a_{eq}}{v_{(0,n)}} \left( f_{(0,n)} - \Delta f \right) \tag{13}$$

where $w$ is the speed of sound and $v_{(0,n)}$ is the zero of the first derivative of the spherical Bessel function of order 0 with twelve digits of precision [29].

Below, we show a summary of the steps followed to assess the contributions to $\Delta f$ and theoretical half-widths $g$.

First, the thermal boundary layer correction has been applied with the three terms due to the matching of the gas and wall temperature (much bigger than the other corrections), the imperfect thermal accommodation at the wall or temperature jump effect, and the penetration of the thermal wave in the shell:

$$\frac{\Delta f_{th}}{f} = \frac{-(\gamma - 1)}{2a_{eq}} \delta_{th} + \frac{(\gamma - 1)}{a_{eq}} l_{th} + \frac{(\gamma - 1)}{2a_{eq}} \delta_{th,w} \frac{\kappa}{\kappa_w} \tag{14}$$

$$\frac{g_{th}}{f} = \frac{(\gamma-1)}{2a_{eq}}\delta_{th} + \frac{(\gamma-1)}{2a_{eq}}\delta_{th,w}\frac{\kappa}{\kappa_w} - \frac{1}{2}(\gamma-1)(2\gamma-1)\left(\frac{\delta_{th}}{a_{eq}}\right)^2$$

with:

$$\delta_{th} = \left(\frac{\kappa}{\pi\rho C_p f}\right)^{1/2} \tag{15}$$

$$\delta_{th,w} = \left(\frac{\kappa_w}{\pi\rho_w C_{p,w} f}\right)^{1/2} \tag{16}$$

$$l_{th} = \frac{\kappa}{p}\left(\frac{\pi M T}{2R}\right)^{1/2}\frac{2-h}{h}\frac{1}{C_v/R + 1/2} \tag{17}$$

where $\gamma$ is the adiabatic coefficient, $\kappa$ and $\kappa_w$ are the thermal conductivity of the gas and the cavity wall, respectively; $M$ is the mean molar mass, $R$ is the gas constant, $C_v$ is the molar isochoric heat capacity, $C_p$ and $C_{p,w}$ are the molar isobaric heat capacities of the gas and the shell wall, respectively, $\rho$ and $\rho_w$ are the densities of the gas and the wall, respectively, and $h$ is the thermal accommodation coefficient. Thermodynamic and transport properties of argon have been taken from the polynomial fits stated in the supplements of [30] because they are the most accurate *ab initio* calculations as declared in [31]. Thermodynamic and transport properties of stainless steel grade 316L have been taken from references [17] and [32]. Second order correction in $g_{th}$ is applied (last term in second equation of (14)) to avoid the possibility of non-physical negative excess half-widths at very low pressures. The effect of the gold-coated wall layer has been neglected in the corrective term of thermal wave penetration into the shell due to the fact that $\delta_{th,w}$, estimated with the gold properties of [33], is between 4 to 6 times larger than the specified layer thickness given by the manufacturer of our cavity. Thermal accommodation coefficient $h$ has been obtained from the literature. The value reported for gold coated 304 SS without treatment and argon is $h = 0.92$ [34]. We considered the effect of replacing this value with $h = 0.85$ [34], which is given for gold coated 304 SS with plasma treatment, in the overall uncertainty of the $R$ measurement and obtained a difference in $R$ of less than 0.7 parts in $10^6$.

Therefore, it had little impact and the assumption of using the value of the literature is acceptable. The determination of $h$ depends upon the gas but do not depend upon the metal [35]. Correction due to the coupling of fluid and shell motion was then applied comparing the model from elastic theory for an isotropic spherical shell of [36] without the radiation term because our shell is surrounded by vacuum, versus the fitting procedure proposed in [37]. This case takes an expression with the same functional form as [36] but with the lower radial symmetric mechanical resonance of the shell (breathing mode), $f_{br}$, being an adjustable parameter. Minimizing the deviation between coefficients $A'_1$ for each mode of the speed of sound regression:

$$\frac{w^2}{A_0} = A'_2 p^2 + A'_1 p + 1 + A'_{-1} p^{-1} \tag{18}$$

after correcting the frequency by shell effect:

$$\frac{\Delta f_{sh}}{f_{(0,n)}} = \frac{kp}{1 - \left(\frac{f_{(0,n)}}{f_{br}}\right)^2}$$

$$k = \frac{5 a_{eq}}{6 t \rho_w w_w^2} \tag{19}$$

This value of $k$ is only valid for an ideal gas with specific heat ratio of 5/3. This gives a value of $f_{br}$ =20500 Hz which lies between the (0,6) and (0,7) mode for our quasi-sphere. $w_w$ is the speed of sound in the wall material [16] and $t$ is the thickness of the cavity. This is what was expected because neither (0,6) nor (0,7) modes were clearly visible in our setup during the measurement process. The shell motion model of [37] is the one that gives the best agreement in our case, in the sense that it yields less dispersion between modes.

Additionally, corrections from the geometrical imperfection of the shell surface have been considered: ducts, microphones and the quasi-spherical shape. The gas inlet duct is composed of two tubes in series, the first of radius $r_0 = 5 \cdot 10^{-4}$ m and length $L_0 = 3.8 \cdot 10^{-2}$ m drilled in the boss of the cavity and the second of radius $r_1 = 7 \cdot 10^{-4}$ m and length $L_1 = 2$ m connected to the manifold system and closed at the end by a valve. The model chosen to estimate the shift in frequency and contribution to half-width of the resonance is the electrical impedance equivalent "T circuits", described in [5]:

$$\frac{\Delta f_0}{f} = Re\left(\frac{i\rho w}{4\pi a_{eq}^2 v_{(0,n)} Z_{in}}\right)$$
$$\frac{g_0}{f} = Im\left(\frac{i\rho w}{4\pi a_{eq}^2 v_{(0,n)} Z_{in}}\right) \qquad (20)$$

where the equations to obtain the value of the inlet acoustic impedance of the inlet gas duct $Z_{in}$ are shown in [5] and [38]. $Z_{in}$ depends on the properties of the gas inside the tube (viscous and thermal penetration length, density, speed of sound and adiabatic coefficient of argon), the dimensions of the tubes and the terminal impedance of the closed-end duct $Z_t \to \infty$.

The microphone correction is due to the presence of the two acoustic transducers. Only the simple frequency-corrective model given in [27] is applied, without taking into account contributions to half-widths:

$$\frac{\Delta f_{tr}}{f} = -\frac{\rho w^2 X_m r_{tr}^2}{2 a_{eq}^3} \qquad (21)$$

where $r_{tr}$ = 1.5 mm is the transducer radius and $X_m$ = 7.1·10$^{-11}$ m/Pa is the compliance per unit area of the transducer, estimated from the technical data given in [6] and chapter 5 of [39].

The shape correction is a second order correction due to the triaxially ellipsoidal shape of the cavity and is applied to the acoustic eigenvalues adding the magnitude [40]:

$$\frac{\Delta v_{ell}}{v_{(0,n)}} = \frac{1}{2} \times \frac{8}{135}(\epsilon_1^2 - \epsilon_1\epsilon_2 + \epsilon_2^2)v_{(0,n)}^2 \qquad (22)$$

where $\varepsilon_1$ and $\varepsilon_2$ are the geometrical parameters determined by microwave resonance and given above.

Finally, the corrected acoustic resonance frequencies are referenced exactly to $T_{TPW}$ = 273.16 K by:

$$f_{(0,n)}(T_{TPW},p) = f_{(0,n)}(T_{exp},p)\left(\frac{T_{TPW}}{T_{exp}}\right)^{1/2}\left(\frac{1+\frac{\beta(T_{TPW})}{RT_{TPW}}p}{1+\frac{\beta(T_{exp})}{RT_{exp}}p}\right)^{1/2} \qquad (23)$$

where $\beta(T)$ is the theoretical second acoustic virial coefficient and converted to the speed of sound squared by using equation (13).

## 3. Results and discussion

The speed of sound squared versus pressure is fitted to a virial type equation:

$$w^2 = A_0 + A_1 p + A_2 p^2 + A_3 p^3 \quad (24)$$

where $A_3 = 1.45 \cdot 10^{-18}$ m$^2$·s$^{-2}$·Pa$^{-3}$ is a fixed parameter taken as the most accurate value in argon [27]. Table 3 shows the fitting parameters of equation (24) and their corresponding uncertainties which were estimated by means of a Monte Carlo approach using $10^5$ iterations. The (0,2) acoustic mode has been neglected from the final fitting because the molar gas constant value computed from it is not in agreement with the values obtained from (0,3), (0,4) and (0,5) modes. The parameter $A_0$ for mode (0,2) was 94757.29 m$^2$·s$^{-2}$. We cannot explain the physical reason why the (0,2) mode is anomalous, but the reason that modes (0,6) and higher have not been used in the calculations is because of the closeness to the shell breathing mode.

Therefore, we obtained an average value of $\bar{A}_0$ = 94755.95 m$^2$ s$^{-2}$ (using (0,3), (0,4) and (0,5) modes) with an experimental standard deviation of the mean, $\sigma(\bar{A}_0)$ = 0.13 m$^2$ s$^{-2}$, this value gives a contribution, due to the dispersion of the modes, to the relative uncertainty of $R$ of 1.4 parts in $10^6$.

Figure 3 shows the normalized residuals for (0,3), (0,4), and (0,5) modes which are less than 1.5 parts in $10^6$. These residuals are smaller than the uncertainty of the parameter $A_0$ (1.8 parts in $10^6$) which has been considered in the final value of the molar gas constant reported in this work.

Figure 4 shows the relative acoustic excess half-widths, i.e. measured half-width $g$ minus the sum of acoustic model estimations from the thermal boundary layer $g_{th}$, duct correction, $g_0$ and classical viscothermal dissipation in the bulk of the fluid, $g_{cl}$:

$$g_{cl} = f^3 \frac{\pi^2}{w^2}\left[\frac{4}{3}\delta_v^2 + (\gamma - 1)\delta_{th}^2\right] \quad (25)$$

with:

$$\delta_v = \left(\frac{\eta}{\pi \rho f}\right)^{1/2} \quad (26)$$

where $\eta$ is the shear viscosity of the fluid, taken for argon from supplement A of [30].

As an estimator of the limits of our acoustic model, we considered the zero pressure limit obtained by fitting the relative excess half-widths, $\Delta g/f$, with a quadratic function of the pressure, i.e. the same function which is used to interpolate the squared acoustic frequencies. The intercept of these fits was -0.1 ppm for mode (0,3), 1.2 ppm for mode (0,4), and 4.3 ppm for mode (0,5) and their contribution to the uncertainty of $R$, as $2\Delta g/f$, amounts to 3.7 ppm.

From the average value of $A_0$ and using:

$$R = \frac{MA_0}{\gamma^{\text{pg}} T_{\text{TPW}}} \tag{27}$$

where $\gamma^{\text{pg}} = 5/3$; $T_{\text{TPW}} = 273.16$ K and argon molar mass $M = 39.947798$ g·mol$^{-1}$ estimated as indicated above, the molar gas constant determined from this work is $R = (8.314449 \pm 0.000059)$ J·K$^{-1}$·mol$^{-1}$. Taking the reference value of the Avogadro constant as $N_A = (6.022140857 \pm 0.000000074) \cdot 10^{23}$ mol$^{-1}$ from CODATA 2014 [21], the derived value of the Boltzmann constant $k_B = R/N_A$ from this work is $k_B = (1.380647 \pm 0.000010)\, 10^{-23}$ J·K$^{-1}$.

Table 4 shows the summary of the uncertainties stated in the above sections of this paper, these uncertainty contributions are related to $R$ as relative standard uncertainties obtaining a total amount of 7 parts in $10^6$. As can be seen, the main contributions to this uncertainty are due to the radius determination using microwave measurements which amounts to 5.1 parts in $10^6$ and the acoustic measurements, 4.4 parts in $10^6$, where the main contribution (3.7 parts in $10^6$) is the relative excess half-widths associated with the limit of our acoustic model. The uncertainty contribution of the molar mass to $R$ is 1.5 parts in $10^6$, and the contributions due to the temperature and pressure measurements amounts to 0.9 parts in $10^6$ and 0.2 parts in $10^6$, respectively.

Figure 5 plots the deviations between the $R$ measurement of this research and the data considered as relevant to the 2014 CODATA adjustment, the most up to date value of $R$ at the moment of writing this paper [21]. These input data are from National Physical Laboratory (NPL) [41], [3], [22], National Institute of Standards and Technology (NIST) [27], [42], Laboratoire National de Métrologie et d'Essais (LNE) [43], [4], [44], Instituto Nazionale di

Ricerca Metrologica (INRiM) [45], [28], National Institute of Metrology (NIM) [23], [42]. Although our uncertainty is greater than the most recent determinations of the molar gas constant, the value reported in this paper lies -1.4 parts in $10^6$ below the recommended value of 2014 CODATA, although still within the range consistent with it.

**Acknowledgements.**

The authors wish to thank to the Spanish Ministry of Finance and Competitiveness for their support through project ENE2013-47812-R and the Regional Government of Castilla y León through project VA035U16.

**Table 1.** Internal equivalent radius and its standard uncertainty[a] of the triaxial ellipsoidal resonance cavity at $T_{\text{TPW}}$ and at the pressures of the acoustic measurements. The results are the average of the microwave $TE_{11}$, $TM_{12}$, $TE_{12}$, $TM_{13}$, $TE_{13}$ modes.

| Pressure/MPa | Radius/m |
|---|---|
| 1.00204 | 0.04003178 |
| 0.90106 | 0.04003175 |
| 0.80111 | 0.04003171 |
| 0.70102 | 0.04003168 |
| 0.60097 | 0.04003164 |
| 0.50049 | 0.04003161 |
| 0.40098 | 0.04003157 |
| 0.30115 | 0.04003154 |
| 0.20100 | 0.04003150 |
| 0.15095 | 0.04003149 |
| 0.10095 | 0.04003147 |
| 0.09092 | 0.04003147 |
| 0.08090 | 0.04003147 |

[a] $u_{\text{radius}}(k=1) = 1.1 \times 10^{-7}$ m

**Table 2.** Internal equivalent radius $a_{eq}$ of the resonance cavity at $T_{TPW}$ extrapolated at zero pressure obtained for different microwave modes.

| Mode | Radius/m |
|------|----------|
| TE11 | 0.04003171 |
| TM12 | 0.04003171 |
| TE12 | 0.04003148 |
| TM13 | 0.04003140 |
| TE13 | 0.04003128 |

**Table 3.** Parameters of the adjustment of the speed of sound to virial type equation (24) and their corresponding uncertainties. Average value of the parameter $A_0$.

| Mode | $A_0$/m² s⁻² | $A_1$/m² s⁻² Pa⁻¹ | $A_2$/m² s⁻² Pa⁻² |
|---|---|---|---|
| (0,3) | 94756.13 ± 0.17 | 2.262·10⁻⁴ ± 1.6·10⁻⁶ | 5.37·10⁻¹¹ ± 2.5·10⁻¹² |
| (0,4) | 94756.03 ± 0.17 | 2.248·10⁻⁴ ± 1.6·10⁻⁶ | 5.26·10⁻¹¹ ± 2.5·10⁻¹² |
| (0,5) | 94755.69 ± 0.17 | 2.169·10⁻⁴ ± 1.6·10⁻⁶ | 5.64·10⁻¹¹ ± 2.5·10⁻¹² |
| $\bar{A}_0$ | 94755.95 m² s⁻² | $\sigma(\bar{A}_0)$ | 0.13 m² s⁻² |

**Table 4.** Uncertainty budget for the determination of the molar gas constant *R*. All the contributions are estimated as relative standard uncertainties.

| Source | | Contribution to R, $10^6 \cdot u_r(R)$ (J·K$^{-1}$)/(J·K$^{-1}$) | Total $10^6 \cdot u_r(R)$ |
|---|---|---|---|
| | *State-point uncertainties* | | |
| Temperature | Calibration | ±0.4 | **±0.9** |
| | Drift | ±0.2 | |
| | Stability | ±0.6 | |
| | Gradient across resonator | ±0.5 | |
| Pressure | Calibration | ±0.14 | **±0.2** |
| | Repeatability | ±0.14 | |
| Molar mass | Isotopic composition | ±1.2 | **±1.5** |
| | Impurities | ±0.8 | |
| | *Cavity radius* | | |
| Microwave radius | Statistical from the intercept of a versus pressure | ±0.2 | **±5.1** |
| | Microwave mode dispersion at p = 0 Pa, $\sigma(\bar{r})$ | ±4.3 | |
| | Relative excess halfwidth | ±2.7 | |
| | *Fitting and corrections to acoustic frequency* | | |
| Acoustic frecuency | Frequency fitting for $A_0$ determination | ±1.8 | **±4.4** |
| | Dispersion of modes, $\sigma(\bar{A}_0)$ | ±1.4 | |
| | Relative excess halfwidth | ±3.7 | |
| | Thermal accomodation coefficient | ±0.7 | |
| **Sum of all contributions to R** | | | **±7.0** |

**Figure 1.** Plot of the triaxial ellipsoidal resonance cavity used in this work: 1. resonance cavity; 2. antenna; 3. electroacoustic transducer; 4. copper block; 5. thermometer (CSPRT), 6. to vacuum, 7. inlet tube; 8. isothermal shield, 9. vacuum jacket. Picture of the quasi-hemisphere with the acoustic transducers.

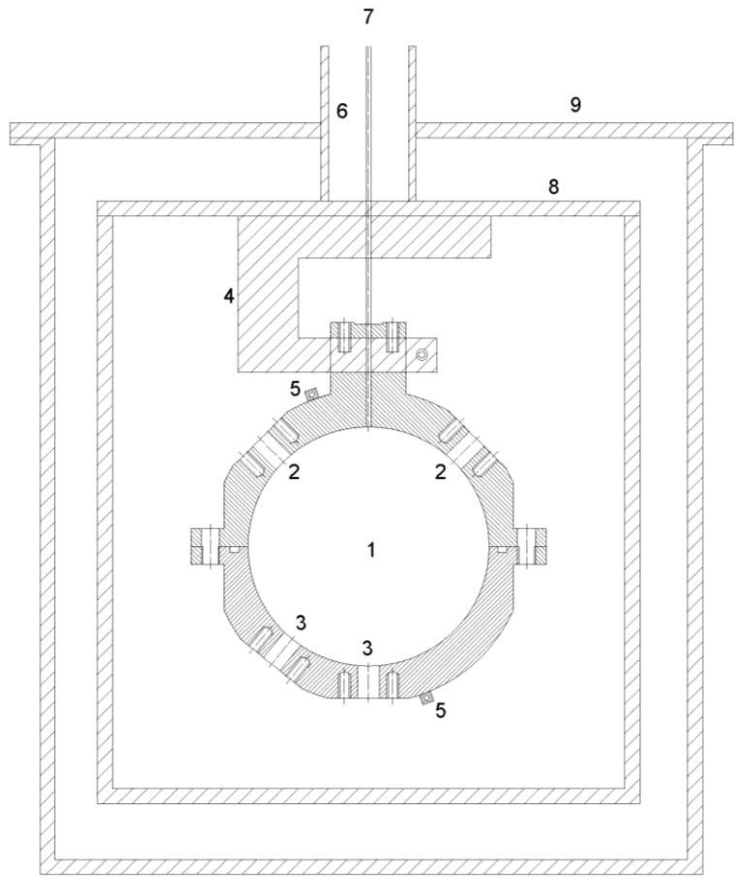
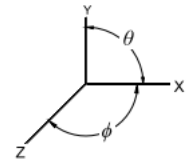
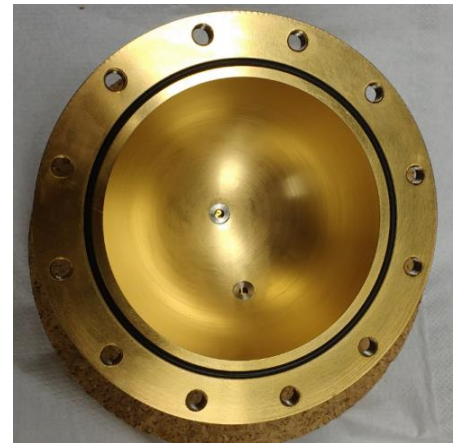

**Figure 2.** Relative microwave excess half-widths for the modes: $TE_{11}$, $TM_{12}$, $TE_{12}$, $TM_{13}$, $TE_{13}$ at $T_{TPW}$ as function of frequency.

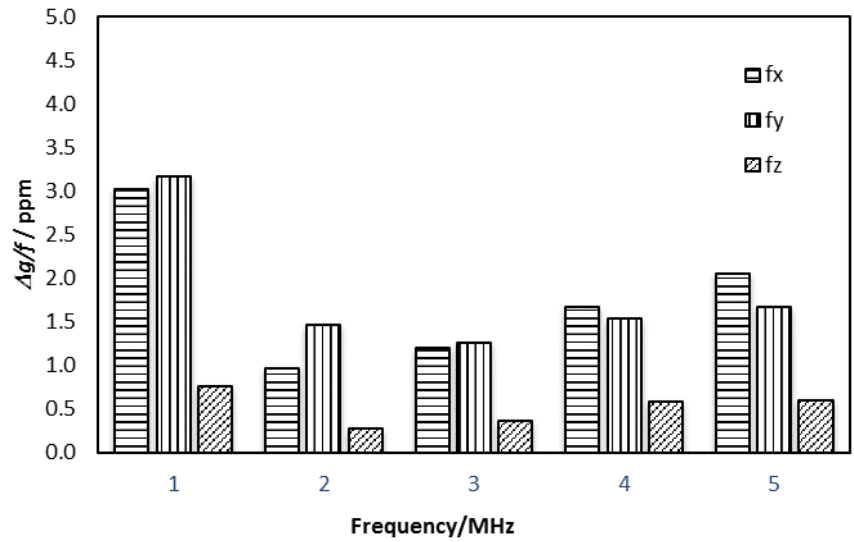

**Figure 3.** Normalized residuals of data fitting to equation (24) for the modes: × (0,3), Δ (0,4), and ◊ (0,5).

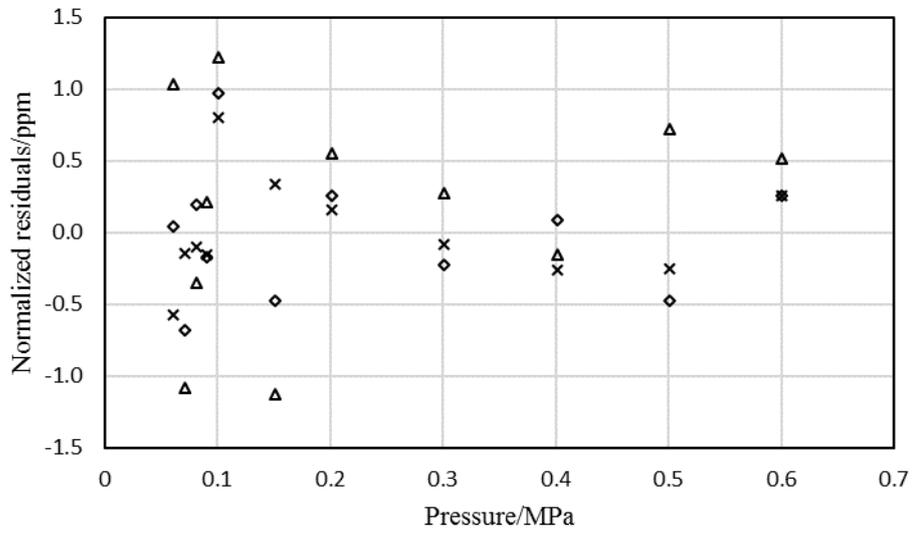

**Figure 4.** Relative acoustic excess half-widths for the modes: □(0,3), ◊ (0,4), and Δ (0,5), at $T_{TPW}$ as function of pressure.

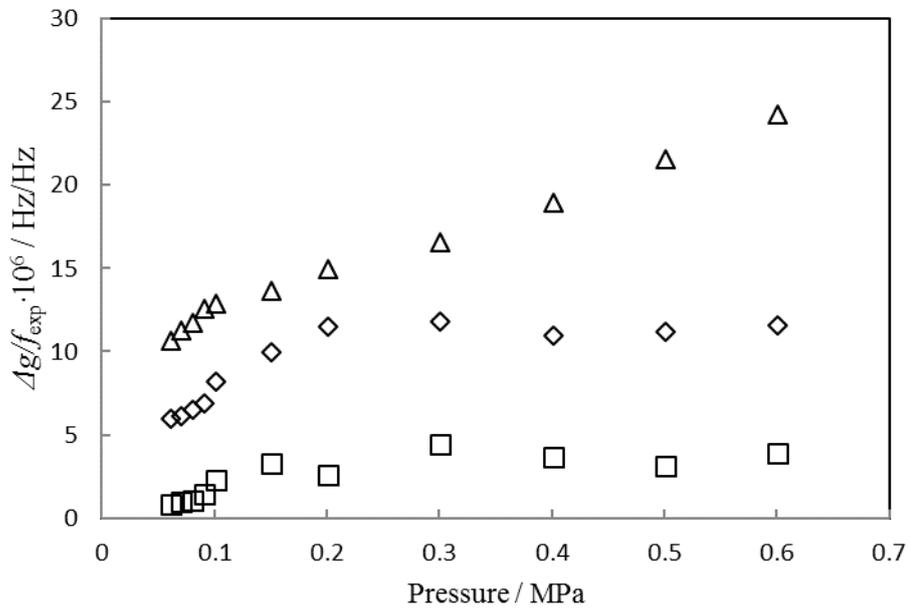

**Figure 5.** Comparison of the determination of *R* from this work with the values considered as input data in 2014 CODATA adjustment. The number to the right of each name indicates the year of publication of each work.

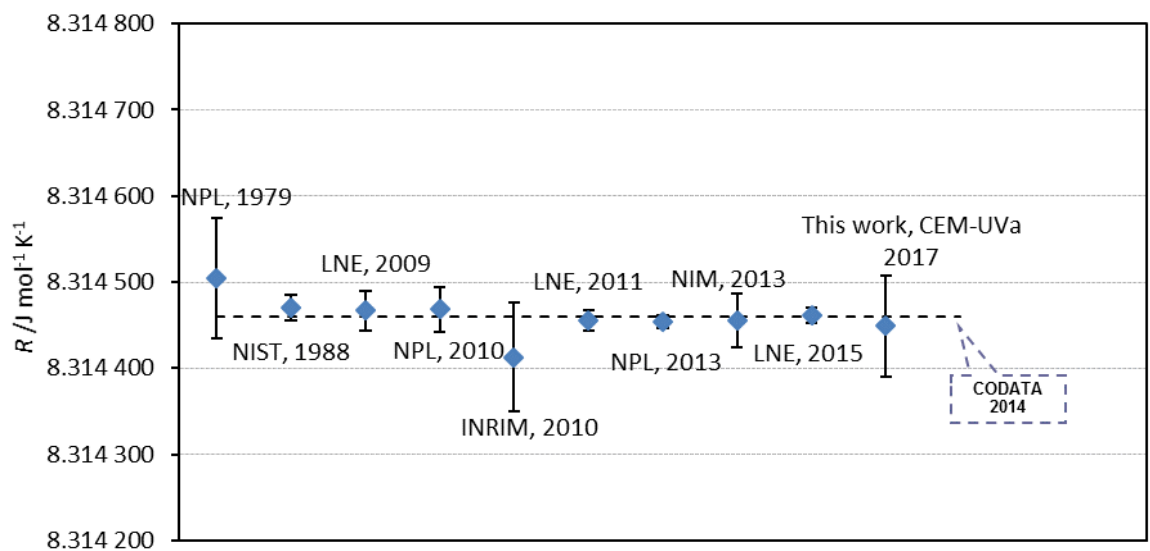